# Prime focus wide-field corrector designs with lossless atmospheric dispersion correction

Will Saunders*[a], Peter Gillingham[a], Greg Smith[a], Steve Kent[b], Peter Doel[c]
[a]Australian Astronomical Observatory, PO Box 915, North Ryde, NSW 1670, Australia
[b]Fermilab, PO Box 500, Batavia, IL 60510, USA
[c]Department of Physics, University College London, Gower Street, London WC1E 6BT, UK

## ABSTRACT

Wide-Field Corrector designs are presented for the Blanco and Mayall telescopes, the CFHT and the AAT. The designs are Terezibh-style, with 5 or 6 lenses, and modest negative optical power. They have 2.2°-3° fields of view, with curved and telecentric focal surfaces suitable for fiber spectroscopy. Some variants also allow wide-field imaging, by changing the last WFC element. Apart from the adaptation of the Terebizh design for spectroscopy, the key feature is a new concept for a 'Compensating Lateral Atmospheric Dispersion Corrector', with two of the lenses being movable laterally by small amounts. This provides excellent atmospheric dispersion correction, without any additional surfaces or absorption. A novel and simple mechanism for providing the required lens motions is proposed, which requires just 3 linear actuators for each of the two moving lenses.

## 1. INTRODUCTION

At the 2012 Astronomical Telescopes and Instrumentation SPIE conference, there were at least four projects proposing large fields of view on 4 meter-class telescopes, for large area multi-object spectroscopic surveys. BigBOSS [1] was proposed for the Mayall telescope, DESpec [2] for the Blanco, HECTOR [3] for the AAT, and 4MOST for VISTA [4]. Since 2012, BigBOSS and DESpec have been merged into the DESI project, and this merger included studies into the best choice of telescope between the Mayall, Blanco and also the CFHT. There are then four classic equatorial telescopes (Mayall, Blanco, CFHT, AAT) of similar size (3.6-4.0m), focal length (12-15m) and F-ratio (f/2.8-f/3.8), looking for a WFC for spectroscopy; designs for each are presented here. 4MOST presents very different challenges, being for the Cassegrain focus on a modern alt-az telescope, and a WFC design for 4MOST is presented separately [5].

We started from the Terebizh WFC designs [6], which are all-silica, allowing superb UV (and NIR) transparency. However, the Terebizh designs are not directly suitable for fiber spectroscopy. Firstly, they have flat but non-telecentric focal surfaces for imaging; for multi-fiber spectroscopy, it is far preferable to have a telecentric[1] focal surface even if it is curved, as pioneered by LAMOST [7], and allowed by both $\theta$-$\phi$ and tilting-spine actuators [1], [8], [9]. Secondly, they do not include an Atmospheric Dispersion Corrector (ADC). An ADC is very desirable for fiber spectroscopy (because very wide wavelength ranges may be observed simultaneously). However, the traditional Amici ADC (two counter-rotating wedged crown/flint doublets) is complex, expensive, and heavy; it also absorbs light just at the blue/UV wavelengths where dispersion correction is most needed, and does so even when at small Zenith Distances where it provides little benefit. Because of these shortcomings, various modifications have been proposed: (a) The WFC design for SuprimeCam on Subaru [10] has a single unpowered flint/crown doublet which is moved laterally; a mechanically and optically much simpler arrangement, but still incurring an extra lens and absorption losses; (b) a design with a single moving weak singlet lens has been proposed for a WFC for the DCT [11], and (c) the final ADC design for BigBOSS consisted of a pair of counter-rotating singlet prisms, with the beam nearly collimated at that point to minimise astigmatism; this provides excellent dispersion correction and image quality, but requires large glass masses and asphericities.

* will@aao.gov.au
[1] Throughout this paper, telecentric is used (incorrectly) to mean that the marginal rays anywhere in the field have no significant excess over the on-axis value. This differs from the standard (but also incorrect) usage that the chief rays are always perpendicular to the focal surface.

We here present a new ‘Compensating Lateral’ ADC. The basic idea is that (a) when a weakly powered spherical lens is displaced laterally, the optical effect is as if a thin prism had been introduced into the beam, with the angle of the prism proportional to the lateral displacement; and (b) if there is another such lens elsewhere in the optical train, it can be displaced in such a way as to compensate for the tilt and astigmatism introduced by the first displacement. The two displacements together give Atmospheric Dispersion Correction of variable strength, with almost no impact on monochromatic image quality. The lenses can be positively or negatively powered, and have any reasonable separation (though more is better). This concept allows a ‘lossless ADC’, in that there are no additional elements or absorption losses. The only penalties are that (a) the ADC setting may need to be frozen during an individual exposure to avoid changing the distortion pattern, and (b) either the two moving lenses must be slightly oversized to accept the full field away from the Zenith, or a small amount of vignetting accepted.

It happens that in the Terebizh WFC design (Figure 1), there are two weak spherical lenses (C3, C4), at large separation, and with no powered elements between them. They thus make a suitable pair for this concept. There are two minor drawbacks, in that (a) C3 is very close to C2, so C2/C3 would normally be built as a doublet, and (b) both lenses are powered in the same (positive) sense, so their displacements must be in opposite directions, ruling out an entirely passive system.

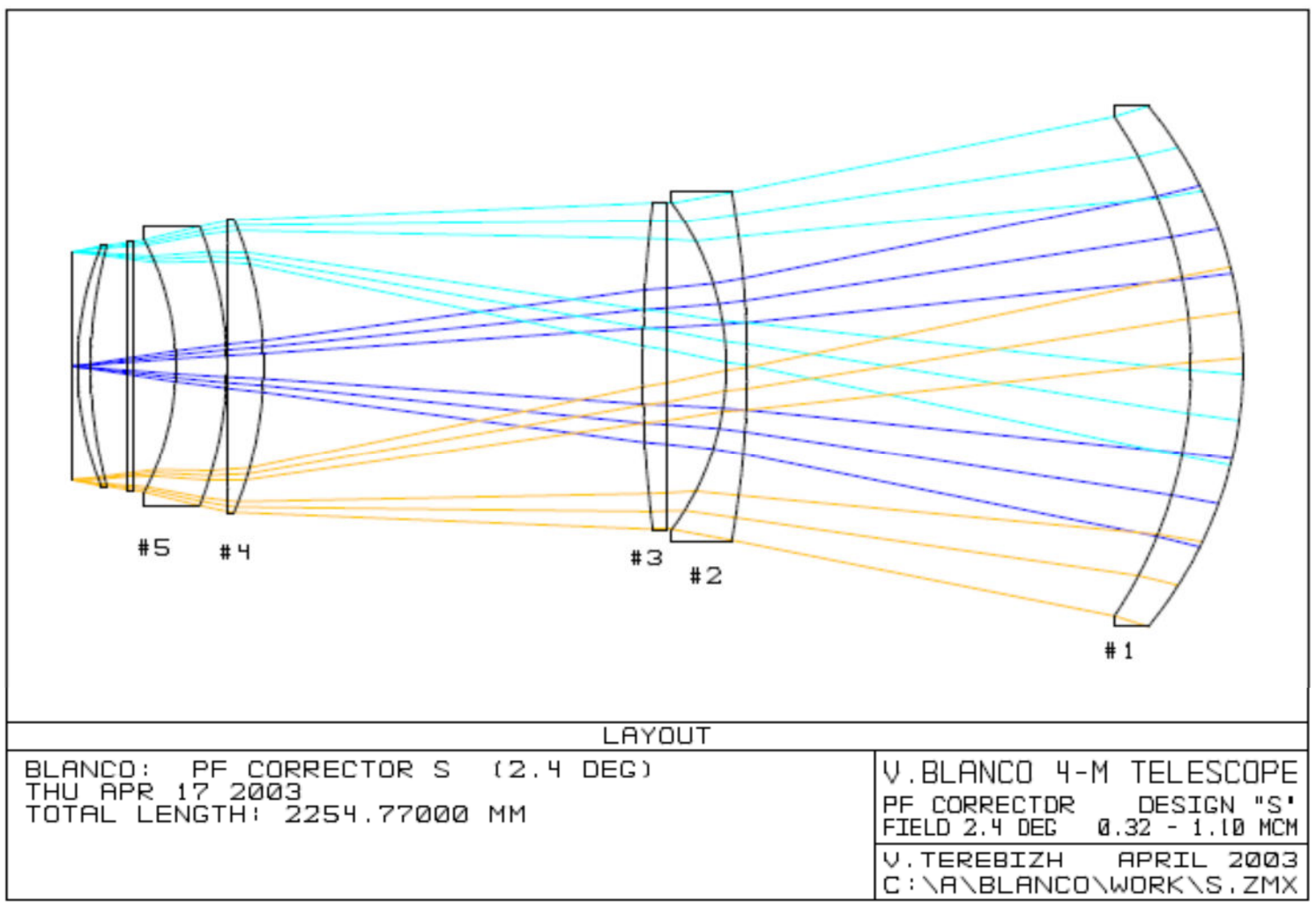

Figure 1. The Terebizh 6-lens design for a 2.4° WFC for the Blanco, for imaging.

The Terebizh design has been adapted and implemented for the Dark Energy Camera (DECam) on the Blanco telescope, including a reduction in lens count to 5 [12]. The DESpec concept was to use C1-C4 of DECam, together with additional lenses, to allow spectroscopy. Unfortunately, the DECam design is unsuitable both for retrofitting an Amici-type ADC (because of image quality degradation), and for implementation of the CLADC concept with the existing lenses, (because C4 is strongly aspheric, and because the lenses are already cemented in place). Designs were sought with additional CLADC-type lenses, but no suitable design was found. For DESpec, the wavelength range of interest for each target was in general rather narrow, and it was found that an ADC provided little overall survey speed benefit. The final DESpec WFC design had no ADC, but had two new lenses replacing the camera window, providing telecentric imaging over a 2.2°, 450mm diameter curved focal surface. The salient parameters of that design (‘DESPEC10’) are included in Table 1 for completeness.

## 2. 2.5° AND 3° DESIGNS FOR DESI

The first CLADC design (MSDESI9_AS2) was an adaptation of the DESpec design, for use on the Mayall telescope (which is slower but otherwise identical to the Blanco) for the DESI project. It allows spectroscopy over a 2.5° FOV, and imaging over 2.2° FOV (using DECam), so the Blanco and the Mayall would then have had interchangeable instruments, allowing both spectroscopic and imaging Dark Energy surveys over the entire extragalactic sky. It was intended to have minimal risk and cost, and to use the same size hexapod as DECam, while still allowing 5000+ fibers in a compact Echidna-style fiber positioner similar to that already proposed for DESpec [13,8]. However, optimising for both imaging and spectroscopy meant some penalty in image quality in both modes, and there was no interest within the DESI consortium for using the Mayall for imaging, but a strong desire for a larger FOV; therefore a 3° FOV version was also developed, for spectroscopy only (MSDESI10_3AS).

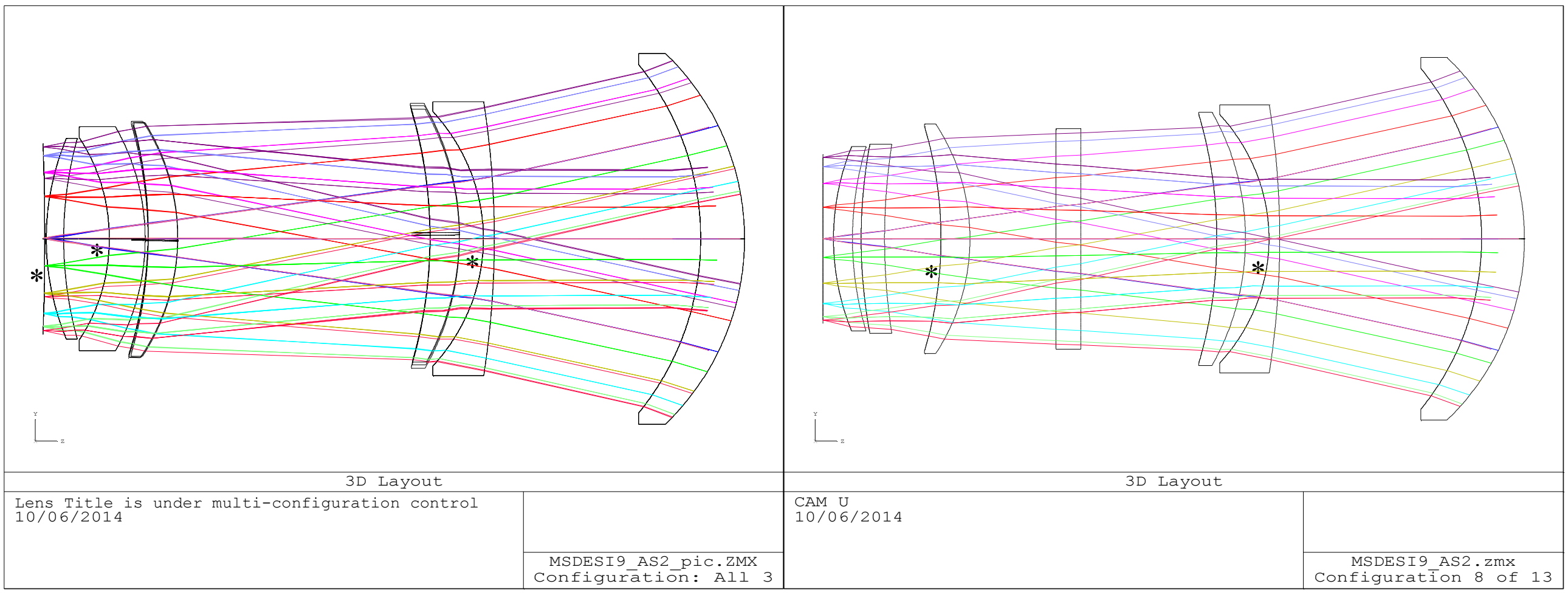

Figure 2. (a) MSDESI9_AS2 design in spectroscopy mode, with 2.5° FOV, for ZD=0, 45°, 60°. The ADC action is provided by lateral movement of C3 and C4, as shown. There are 2 aspheric glass surfaces, as shown, and the focal surface is aspheric also. (b) MSDESI9_AS2 design in imaging mode, with 2.2° FOV and using DECam. C6 is the existing DECam camera window; if a new camera was built, the last two lenses would be combined.

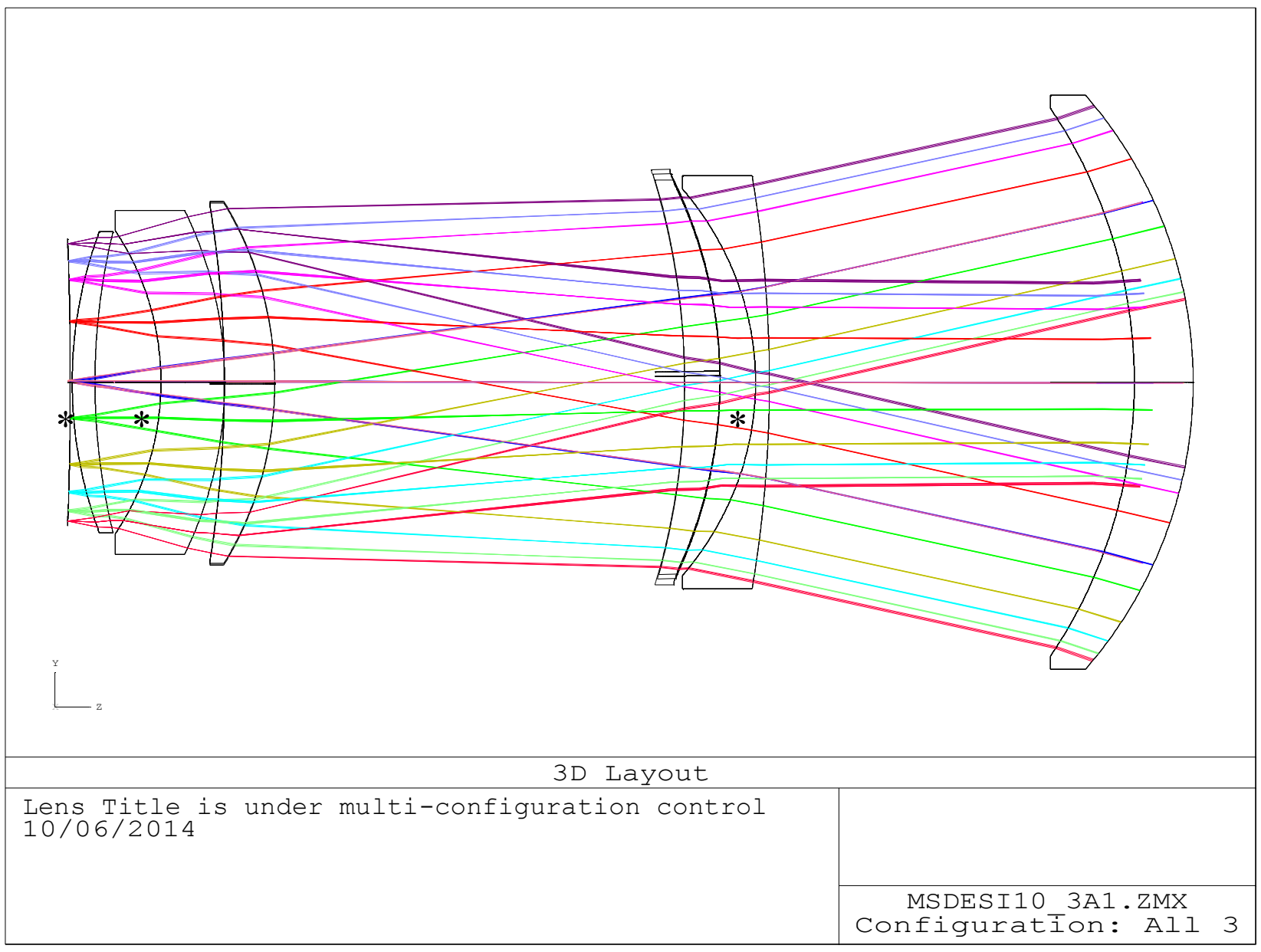

Figure 3 MSDESI10_3A1 design with 3° FOV, for ZD=0, 45°, 60°. To the same scale as Figure 2. Aspheres as marked.

The DESI WFC design was selected in July 2013, while the positioner technology selection still remained open. Therefore, a programmatic decision was made that the WFC design should not constrain the positioner selection, which meant enforcing a WFC speed of f/3.6 or slower, to accommodate the larger pitch of the competing $\theta$-$\phi$ positioner designs. This ruled out the designs presented here, and they are presented in the hope they will be useful for other projects or telescopes. The design selected for DESI is detailed elsewhere [14].

For both the Mayall designs, fused silica was used throughput, because of its lightness and stiffness-to-weight ratio, its UV transmission, and the existing experience of making such lenses for DECam. The choice of which surfaces to aspherize was driven by the desire to only aspherize concave surfaces (for ease of testing), and the preference to keep C3 and C4 spherical to allow them to function better as an ADC. The final choices were the second surface of C2, and for spectroscopy the second surface of C5, and for imaging the first surface of C5. Allowing the focal surface to also be aspheric gave significant benefits to the image quality, the telecentricity, and the tolerance to defocus. Such a surface can be straightforwardly be achieved for Echidna-style tilting spines; either they can be mounted on an aspheric reference surface [AS14], or they can be individually adjusted for length prior to assembly.

The designs were optimised for 350-1050nm, and at ZD=0, 45°and 60°. The clear diameters for C3 and C4 were set by their unvignetted value ZD= 45°[2]. Marginal ray angles in excess of the on-axis value were penalised (this is more tolerant than setting the telecentricity from the chief rays, because the beam becomes slower off-axis). The ADC action is achieved by moving C3 up by ~11mm and C4 down by ~4mm at ZD=45°, and 19mm/6mm at ZD=60°, combined in each case with a tilt. Early designs quickly established that for best performance, the lenses should 'follow the curve' as they move laterally (Figure 2)[3]. The motion was constrained to be about a fixed horizontal axis of rotation, orthogonal to the telescope axis. For an equatorial telescope, the telescope rotates with respect to the gravity vector, so this horizontal axis is not fixed with respect to telescope. A mechanism for providing the required motions is proposed in Section 7.

The maximum rates of change of deviation from spherical are 18µm/mm for C2, and 8µm/mm for C5. The focal surface shape for spectroscopy is elliptical, with maximum deviations from a sphere ±250µm. This was checked back against the designs for DESpec, to check that such a positioner could also be used one day on the Blanco, albeit with new C5 and C6 lenses.

The image quality for both designs is ~15.5µm rms and polychromatic (350-1050nm) $r_{80}$<18µm (0.31") at Zenith (Figure 4). The quality of the ADC is very good, with the maximum spot sizes increasing only to 15.9µm at ZD=45° and 17.8µm at ZD=60° (Figure 4). The maximum telecentricity error, using the excess in the marginal ray angle over its value at the center of the field, is just 0.09° at Zenith, 0.15° at ZD=45° and at ZD=60°. For the chief rays, the maximum non-telecentricity is 0.14°, 0.19° and 0.26° respectively.

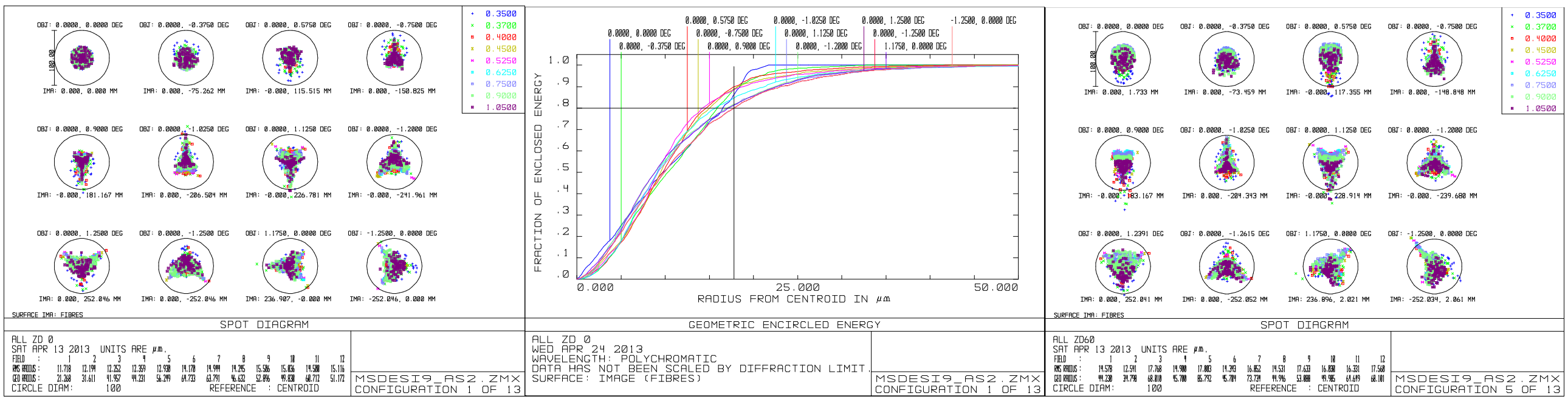

Figure 4. (a) Spot diagrams for MSDESI9_AS2 for 350-1050nm spectroscopy at Zenith. Field-of-view is 2.5°. Circle is 100µm diameter (1.76"). (b) The equivalent geometric encircled polychromatic energy vs radius, also at zenith. The lines show 80% enclosed energy, at 0.313". (c) Spot diagrams for ZD=60°.

For imaging over 2.2° (Figure 5) this design is not as good as DECam, giving $r_{80}$<25µm or 0.43" in *ugrizY*. The reason is that the simultaneous optimisation for imaging and spectroscopy inevitably means compromises for both. However, the ADC is extremely useful for imaging as well as spectroscopy - Figure 5(b) shows the image quality in *u*-band at ZD=60°, without and with the ADC.

2 This causes 1.5% vignetting at ZD=60°, but has the benefit of vignetting the marginal rays with the greatest angles.
3 This had already been discovered independently [DCT].

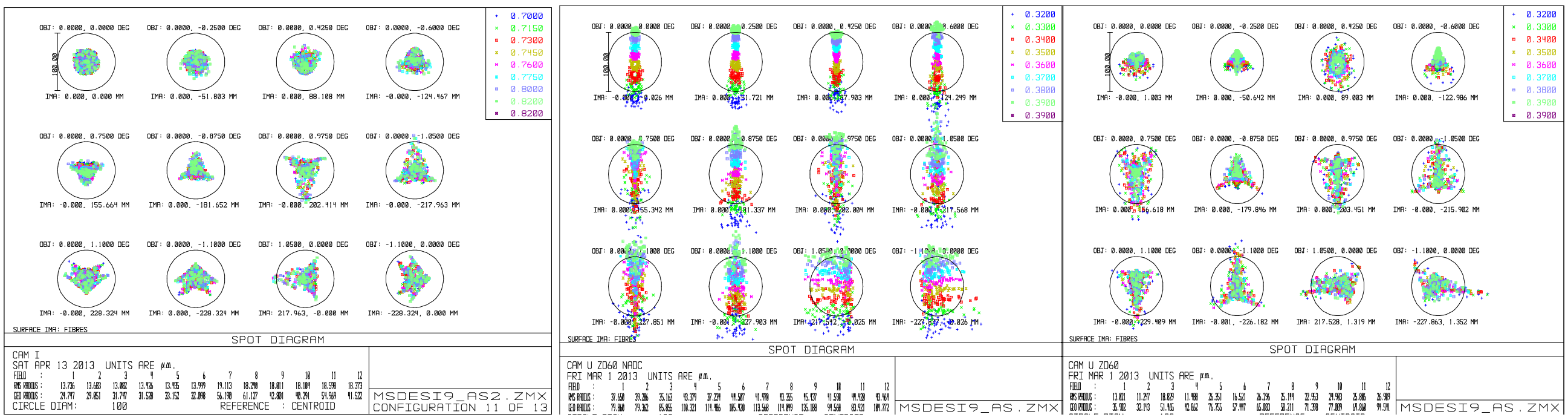

Figure 5. (a) MSDESI9_AS spot sizes for imaging mode, for *i*-band, at Zenith. (b) Imaging quality in *u*-band at ZD=60°, without and with the ADC correction.

## 3. FIELD DISTORTION

When large powered elements are moved laterally in a lens design, there will be significant changes in the field distortion pattern. When these changes amount to a significant fraction of the seeing or the fiber diameter, there will be a loss of efficiency. For the MSDESI9_AS2 design, the image of a target on the telescope axis image moves by ~1.5mm when the ADC setting is changed from zenith to ZD=60°, but this is just an offset to be included in the telescope pointing model. But also, changes of several hundred µm occur in the *relative* positions of images across the focal surface between Zenith and ZD=60°, and this is in addition to (and larger than) the distortion caused by differential refraction. Figure 6(a) shows the fractional distortion of the field in the *X* (azimuthal) and *Y* (altitude) directions, for ZD=45° and ZD=60°, and also shows the difference between the X and Y distortion. The effect is very well described as the change in the fractional *Y* distortion proportional to *Y tan*(ZD), with the *X* distortion unchanged. Figure 6(b) shows the worst case change in the 2D distortion map over 15 minutes at air mass ~1.4, when the ADC is continually adjusted to its optimal value. The changes amount to tens of µm, several times larger than the usual rotation and shearing due to changes in differential refraction. It seems too large to accommodate, unless a positioner with very reliable open-loop motion is available. However, the problem can be avoided simply by setting the ADC to an average value for each frame, and freezing it for the duration of the frame. This introduces an error (in both magnitude and direction) in the dispersion correction at the start and end of the frame, but this is small for any normal air mass and frame length. A worst-case effect on the actual images is shown in Figure 7.

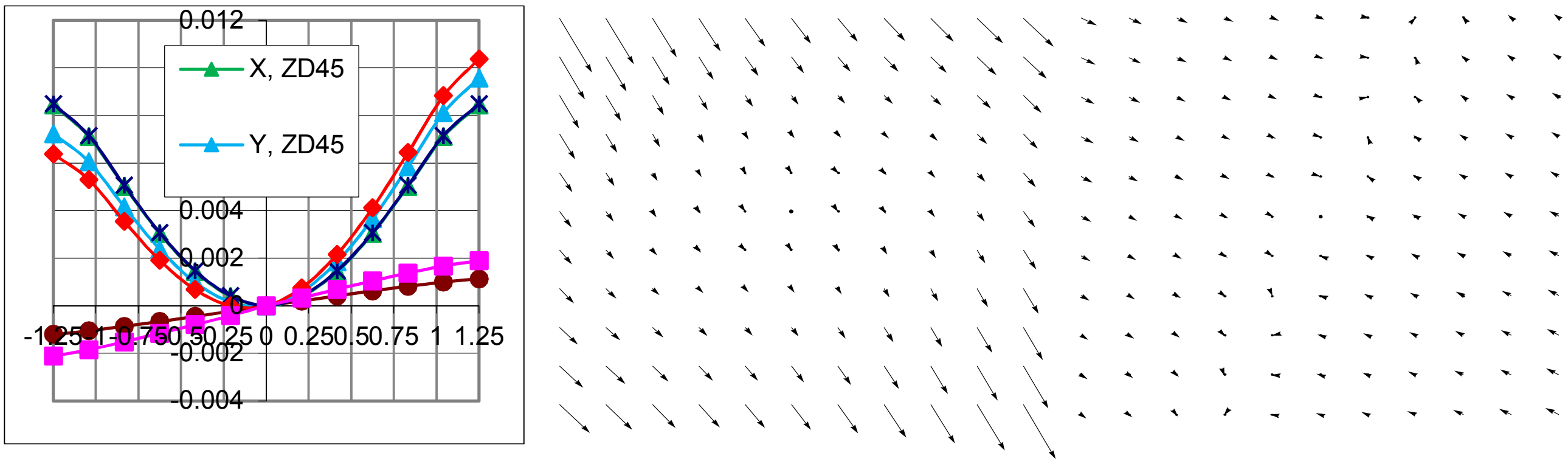

Figure 6. (a) Field distortion for MSDESI9_AS2, for ZD=45° and ZD=60°. The *X* distortion is virtually constant, while the *Y* distortion varies by an amount proportional to *Y tan*(ZD). (b) The resulting change in the distortion pattern arising from adjustment of the CLADC during an 15 minute exposure, for the worst case at ZD=45° ($\delta$=20°, air mass changing from 1.4 to 1.5). The field is 2.5° square, and the vector lengths are scaled by a factor 1000. (c) For comparison, the distortion due to differential refraction for the same field, over the same period, and in the same units.

The change in the distortion depends sensitively on the separation of the two lenses being used as an ADC (and to a lesser extent on the glass choices), varying by almost an order of magnitude between all the different designs presented in this paper. The 3DF design (Section 5) has distortions small enough not to require freezing the ADC while integrating, and this reopens the possibility that an entirely passive ADC could be devised, as has been done for 4MOST on VISTA [5]. Since large distortion changes are not intrinsic to the CLADC concept, any future designs would obviously benefit from making an explicit effort to minimise them, but this was not done for the designs presented here.

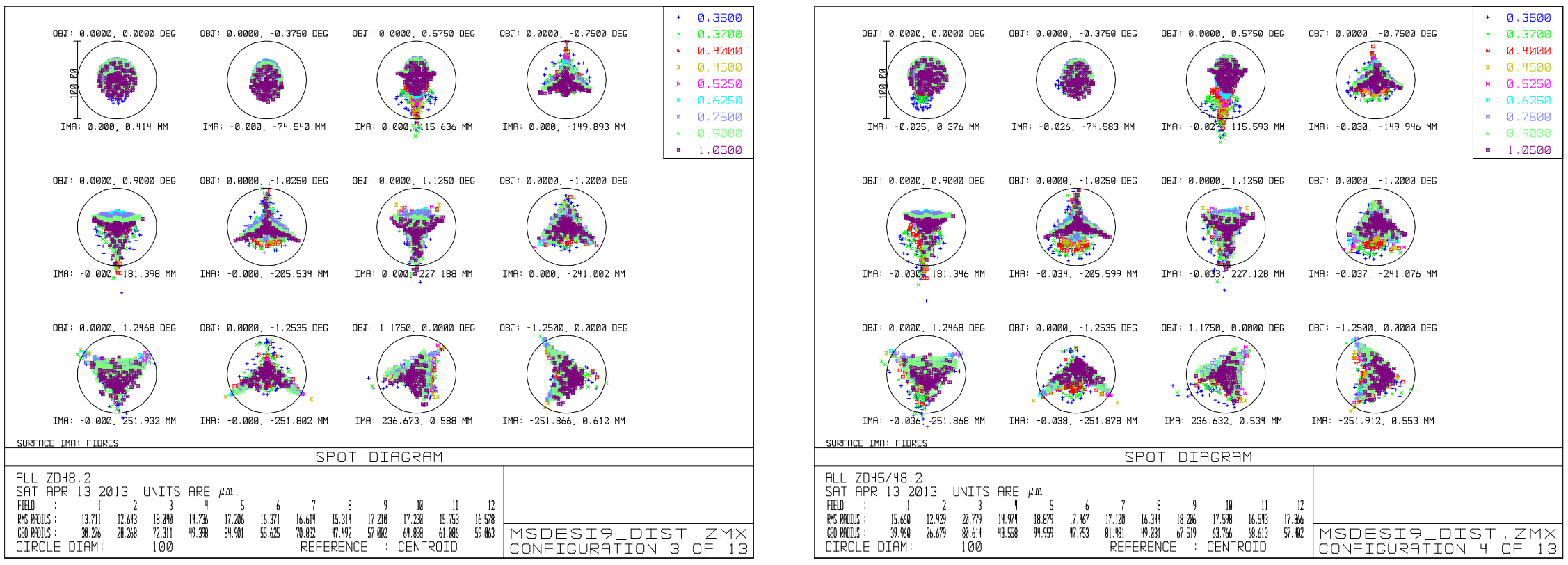

Figure 7. The effect of freezing the ADC setting. (a) Spot diagram for ZD=48.2° (airmass=1.5) with the ADC at the correct setting; (b) spot diagram for ZD=48.2°, with ADC set for ZD=45°, corresponding to the worst case for this ZD and a time error of 15 minutes.

## 4. A 2.5° DESIGN FOR THE CFHT

One telescope option for the DESI project was the CFHT. Although smaller than the Mayall or Blanco telescopes (at 3.58m), the superb site meant it would have had the fastest survey speed (for equal FOV, fiber numbers etc), as well as being able to survey ~10% more sky because of its tropical latitude (18°N) [15]. The 44% gain in survey speed offered by the CFHT over the Mayall [15] means a 2.5° FOV and ~3500 fibers would have sufficed to carry out the DESI survey in the required timescale, and the slower speed allows either $\theta$-$\phi$ or Echidna actuators . Therefore, it was interesting to see if the CLADC concept could be adapted to the CFHT, which is much slower than the Mayall or Blanco (f/3.77 native speed), and a classic Cassegrain rather than a Ritchey-Chrétien.

Optically, this presented no problems, and indeed allowed a simplification to just 5 lenses of fused silica, with one of the two aspheres being very mild. Figure 11 shows the optical layout for 2.5° FOV for spectroscopy. The speed is f/4.14, the focal plane diameter is 662mm, and the glass mass is 653kg. At ZD=60°, the required movement of C4 is ~2mm, and at right-angles to the axis; this means a very simple mounting and translation mechanism (e.g. the lens cell held by 2 actuators and 2 springs) is likely to be feasible.

The performance is somewhat better than the Mayall designs (in angular terms), with polychromatic $r80<0.255''$ at Zenith. Howver, the ADC performance is somewhat worse, because C4 is now such a thick and strong lens.
There remains a question of whether such a large Prime Focus WFC will fit into the CFHT dome, which has much less clearance round the telescope than at the Mayall or Blanco. The CFHT observatory manual [16] gives a distance of the nearest obstruction from the M1 vertex of 15253.5mm. The total track of CFHT_AS2S_15 is 14094.2mm, leaving ~1150mm for fiber actuators, fiber routeing and clearances. This is much more than allowed for e.g. the AESOP positioner design for 4MOST, so should be more than adequate. The total mass should also be fine, since the glass mass of 653kg is a very small fraction of the top end mass limit (which is at least 5800kg).

It was found that a simple 4-lens variant, with the last two lenses combined and only one lens (C3) moving, gave reasonable performance. A 5-lens version with 2.2° FOV and flat focal plane for imaging was also developed, also with 5 lenses and a single moving lens. The image quality is superb (<15µm rms in all bands including $u$, and typically <10µm). The design is available upon request

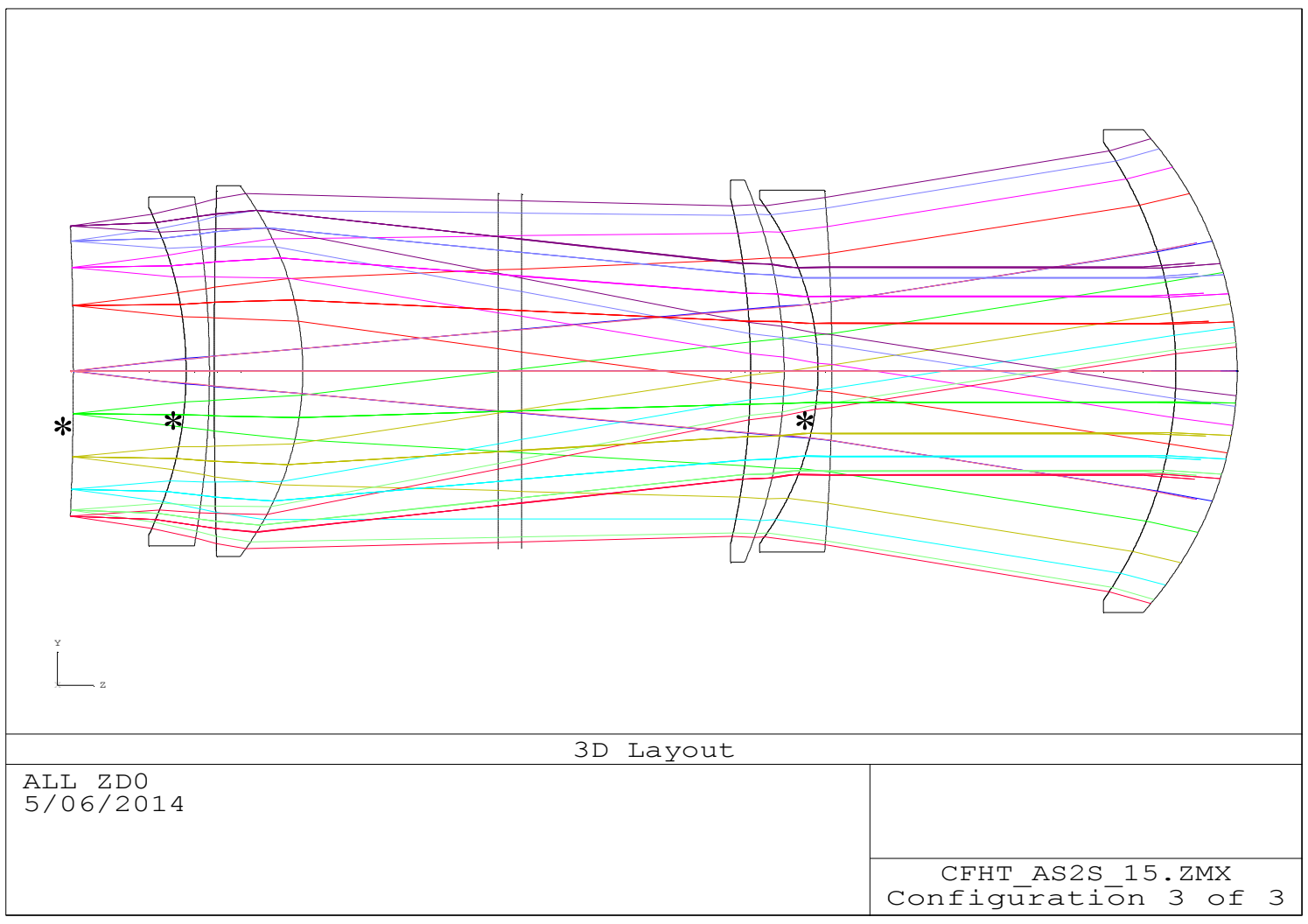

Figure 8. CFHT_AS2_15 design in spectroscopic mode with 2.5° FOV. Aspheres are as marked. ADC lenses are C3 and C4.

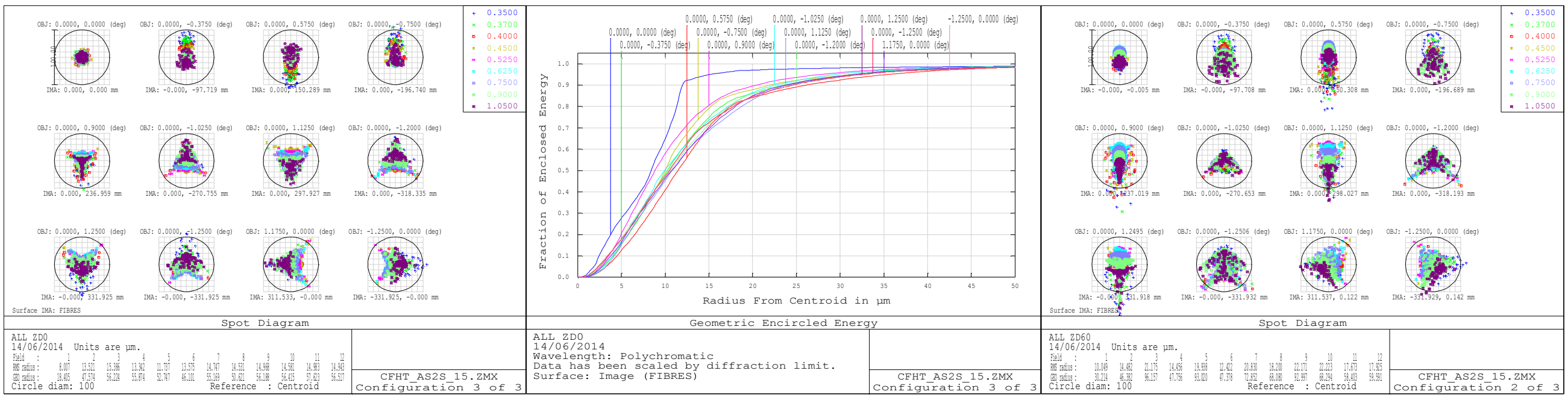

Figure 96 (a) Spot diagram for CFHT_AS2S_15 at Zenith, (b) polychromatic (350-1050nm) enclosed energy, with $r_{80}$ < 0.255″, (c) spot diagram at ZD=60. Circle is 100µm (1.37″).

## 5. A 3° DESIGN FOR THE AAT

An AAO project (HECTOR) is planned for feeding spectrographs from robotically positioned multi-fiber integral field units (IFUs) at the AAT prime focus. Desirably, the field should be large enough to warrant fielding ~100 such IFUs. Although it is conceivable that the IFU feed could share the existing 2dF top end (with its multiple single fiber feeds), there are clear advantages in having an independent top end with a larger field and giving better optical performance than the 2dF corrector. The use with IFUs suggested that an ADC might not be necessary, since post-processing of the spectra could correct for the image shift as a function of wavelength. However, with the diameters proposed for the IFUs, there would be a significant loss of spectrally corrected field in the direction of atmospheric dispersion. So, when it became clear that an ADC could be incorporated with no additional absorption of light and few other complications, the initial 3dF corrector design was modified to include an ADC. The priorities were rather different to the other designs presented in this paper: cost is a strong driver, so a simplified design was sought and some vignetting at the field edges allowed, and the design wavelength range is 400-900nm, so there is no strong reason to keep an all-silica design. Figure 10 shows the layout of the corrector with the elements all aligned axially, as for observing at the zenith, and with the elements displaced (with exaggeration by a factor 10) for observing at ZD 60°.

The specification for imaging onto the IFUs is 80% energy enclosed within 1 arcsec diameter for all individual wavelengths from 400 to 900 nm (with a single compromise focus setting). The design readily meets this at the zenith and the diameter enclosing 80% at ZD 60° increases to barely 1.1 arcsec for some wavelengths. With a spherical plano-convex lens (which would act as the CCD dewar window) replacing the final meniscus, the corrector gives direct imaging across 2° diameter on a flat surface. For each of the spectral windows *g*, *r*, *i*, and z, 90% energy is enclosed within 1 arcsec diameter for ZD 0 to 60°.

Whereas, with some of the CLADC designs, the unsymmetrical distortion requires that the ADC correction be frozen during an exposure of 20 min or more, to avoid image trails longer than those unavoidably associated with changing atmospheric distortion, this effect is small enough with this corrector design to render this unnecessary (the relative image positions across the entire focal plane change by less than 75μm, between zenith and ZD=60°).

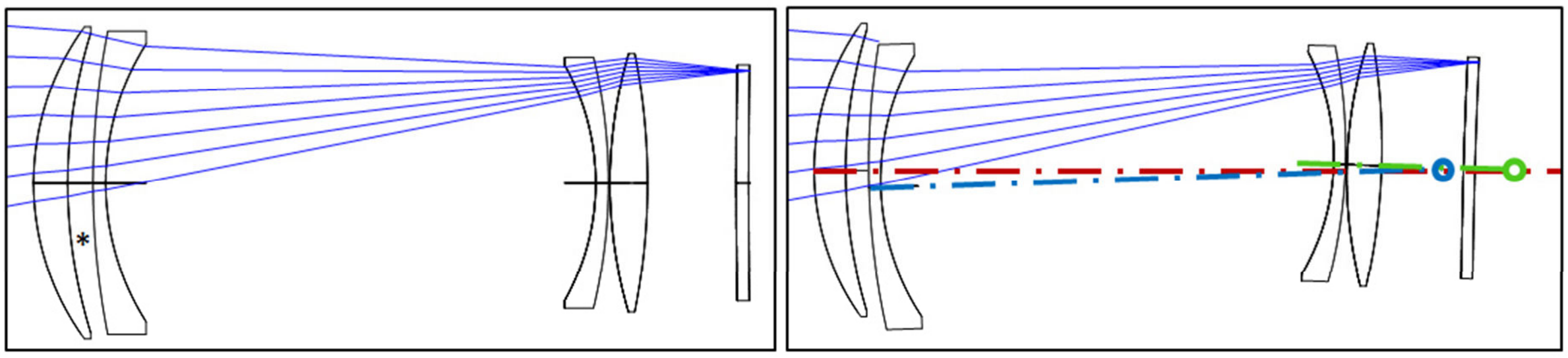

Figure 10. Layout of lens design for feeding IFUs over the 3 degree field diameter, illustrating the ADC action. All surfaces are spherical except that identified with *, which is an ellipsoid. The element materials are, from the left: silica, silica, BSL7Y, silica, and BSL7Y. The outside diameter of the first element is 1000 mm. The robotic IFUs attach to the rear surface of the last element. The exit pupil is concentric with this surface to give efficient coupling into the optical fibers.
(a) At zenith with all elements set on axis. (b) The displacements of the ADC elements for ZD 60° are exaggerated by a factor 10 to make their offsets more obvious. Element 2 is tilted downwards while elements 3 through 5 are tilted upward as a group. The approximate locations of the axes for the tilts are identified by the circles. For ZD 60°, the offsets of the vertices of C2 and C3 are 6.3 mm and 2.3 mm respectively.

## 6. DESIGN SUMMARY

Table 1 gives the salient parameters for each of the designs presented here.

Table 1. Parameters for various designs. Notes to Table 1: All quoted C1 sizes include a 4% margin for mounting, over the unvignetted clear aperture. Attempts to shrink C1 much below its ZEMAX preferred size always caused rapid loss of image quality. The dependence of image quality on field size was also very strong, and this means the gain in survey speed from a larger field-of-view is only about half of what the direct AΩ product would suggest.
* No increase in marginal ray angle, but up to 16% vignetting accepted at edge of field.
** Single asphere is ccv ellipsoid.

| Name MSDESI+ | DESPEC10 | DESI9_AS2 | DESI10_3A1 | CFHT15_AS2 | AAT 3dF |
|---|---|---|---|---|---|
| FOV (°) | 2.5+2.2 | 2.5 | 3 | 2.5 | 3 |
| WFNO | 2.96 | 3.03 | 3.01 | 4.15 | 3.42 |
| EFFL | 11609 | 11489 | 11425 | 14887 | 13342 |
| Max rms radius, 350-1050nm, ZD0 (μm) | 16.6 | 14.7 | 14.1 | 14.9 | 28 |
| Max rms radius, ZD45 (μm) | - | 16.5 | 15.2 | 17.6 | 28 |
| Max rms radius, ZD60 (μm) | - | 16.8 | 18.5 | 21.1 | 29 |
| $r_{80}$, Zenith (″) | 0.336 | 0.319 | 0.308 | 0.255 | 0.5 |
| Max excess for marginal rays (°) | 0.04 | 0.14 | 0.07 | 0.08 | 0* |
| C1 diameter (mm, 4% rim) | 980 | 1020 | 1225 | 1113 | 1000 |
| Lenses (aspheres) | 2 (1) new | 6 (2) | 6 (2) | 5 (2) | 5 (1) |
| Max aspheric departure μm/mm | 21 | 21+10/5 | 22+7 | 1+8 | 5** |
| Glass | Silica | Silica | Silica | Silica | Silica/ BSL7Y |
| Glass mass (kg) | 376 | 439 | 754 | 653 | 489 |

The cost of the WFC will depend very strongly on the glass mass, both because of the direct costs of the blanks and the polishing, and the difficulty of mounting them at prime focus, within the overall mass and flexure allowances. The glass masses have been calculated in ZEMAX, assuming all lenses are 4% oversize, for mounting. For comparison, the DECam glass mass is 382kg. Minimum lens thicknesses were based on the ratio of thickness to diameter. The average thickness (i.e. the volume/area) is always greater than D/12.5. With FEA analysis and dialog with vendors, many of the lenses could surely be lightened, in some cases with beneficial effect on the imaging as well as weight and cost.

## 7. MECHANISING THE ADC MOTIONS

With the requirement to tilt the ADC elements around horizontal axes, the mechanisms to achieve this would be simple if the telescopes were on alt-azimuth mountings. Polar mountings make the issue is more complicated, since the horizontal axes change in their relationship to the corrector structure . The basic mechanism envisaged for the task is illustrated diagrammatically in Figure 11. The cell of the element to be rotated around the appropriate horizontal axis is attached to the main structure of the corrector through three struts of equal length with their axes, when the ADC element is on axis, pointing to the desired center of motion. Three encoded linear actuators attach to the ADC cell at the same three points and these provide for tilting the element about the required axis and controlling its location in Z. There is no over-constraint so, as long as the actuators are appropriately limited in their strokes, the mechanism cannot jam. While the kinematics of the assembly are not perfect, since the three strut axes no longer meet at a point when the element is displaced from its central location, the resulting tilt errors are negligible (0.01°) in relation to the optical tolerances for the small displacements needed.

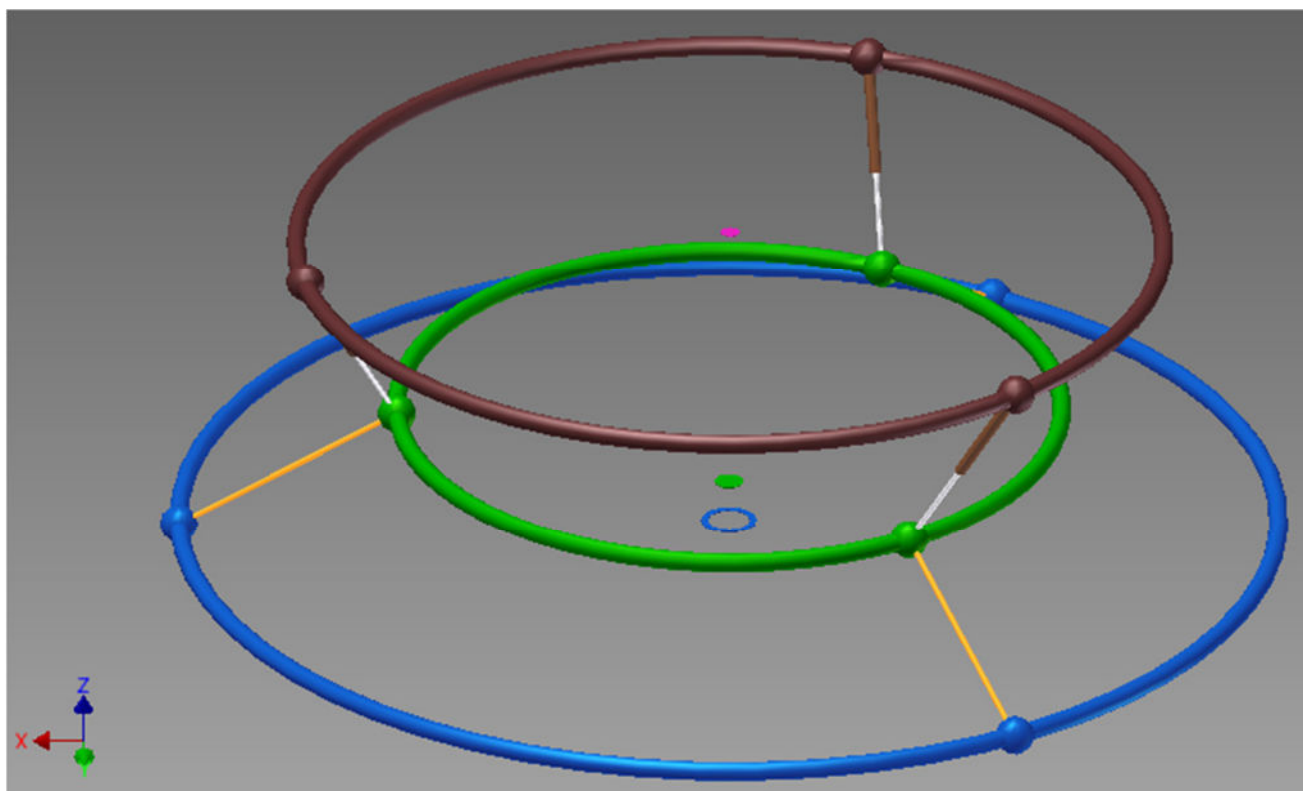

Figure11. Essentials of the mechanism to tilt a lens element as required for the ADC action. The large blue and brown rings are fixed in relation to the main lens structure. The smaller green ring represents the lens element. The yellow members are struts of fixed length which point toward the required centre of rotation. The brown and white members are actuators with absolute linear encoding.

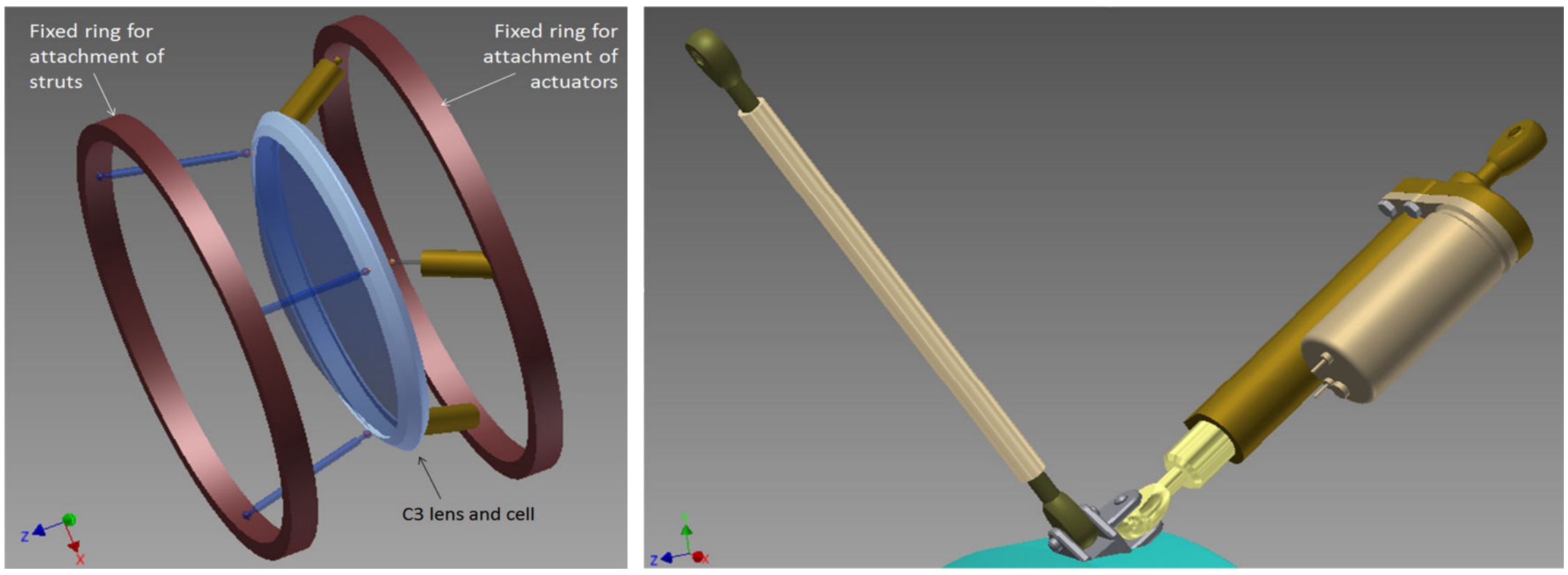

Figure 12. (a) A properly scaled example of the mechanism for controlling tilt of one of the ADC elements, but with the struts and actuators shown only diagrammatically. (b): Strut and actuator shown with realistic modelling, having due regard for the stiffness and drive force requirements.

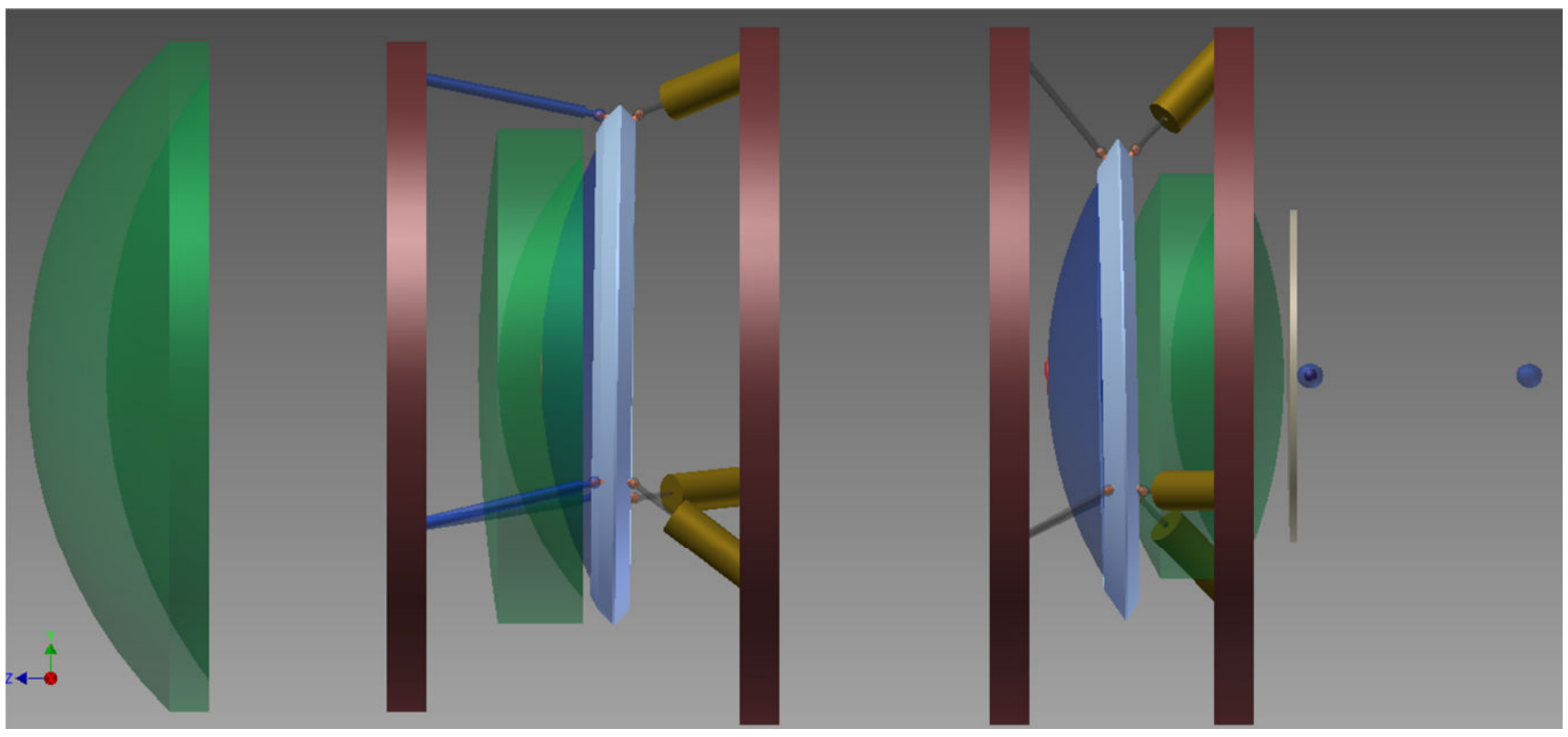

Figure 13. The full set of elements and mechanisms. The ADC elements are shown with greater tilts than required for ZD 60°. The blue spheres show the axial position of the centers of rotation (left for C3, right for C4). The outer support structure need be no larger than already needed to accommodate the C1 cell.